\documentclass[usenatbib]{mn2e}
\usepackage{natbib,psfig}
\usepackage{amsmath}
\catcode`\@=11
\def\gsim{\ifmmode{\,\mathrel{\mathpalette\@versim>\,}}
    \else{$\,\mathrel{\mathpalette\@versim>}\,$}\fi}
\def\lsim{\ifmmode{\,\mathrel{\mathpalette\@versim<\,}}
    \else{$\,\mathrel{\mathpalette\@versim<}\,$}\fi}
\def\@versim#1#2{\lower 2.9truept \vbox{\baselineskip 0pt \lineskip
    0.5truept \ialign{$\m@th#1\hfil##\hfil$\crcr#2\crcr\sim\crcr}}}
\catcode`\@=12  

\newcommand{\Myr}{\,{\rm Myr}}
\newcommand{\Gyr}{\,{\rm Gyr}}

\newcommand{\mnras}{MNRAS}
\newcommand{\aap}{A\&A}
\newcommand{\apj}{ApJ}

\newcommand{\aj}{AJ}
\newcommand{\nat}{Nature}

\def\av#1{\langle#1\rangle}
\def\rms{\rm rms}

\def\fracj#1#2{{\textstyle{#1\over#2}}}

\title[Radio-loud flares from microquasars]
{Radio-loud flares from microquasars and radio-loudness of quasars}

\author[C. Nipoti, K.M. Blundell \& J. Binney]
{Carlo Nipoti$^{1,2}$, Katherine M.\ Blundell$^1$ 
and James Binney$^1$
\\
$^1$
University of Oxford, Department of Physics, Keble
  Road, Oxford, OX1 3RH, U.K.\\
$^2$
Dipartimento di Astronomia, Universit\`a di Bologna,
      via Ranzani 1, 40127 Bologna, Italy\\}

\begin{document}
\date{Accepted 2005 May 12.  Received 2005 April 19; in original form 2005 March 4.}

\pagerange{\pageref{firstpage}--\pageref{lastpage}} \pubyear{2005}

\maketitle

\label{firstpage}

\begin{abstract}
The low-frequency power spectra of the X-ray and radio emission from four
microquasars suggest that two distinct modes of energy output are at work:
(i) the `coupled' mode in which the X-ray and radio luminosities are closely
coupled and vary only weakly, and (ii) the `flaring' mode, which
dramatically boosts the radio luminosity but makes no impact on the X-ray
luminosity.  The systems are in the flaring mode only a few percent of the
time. However, flares completely dominate the power spectrum of radio
emission, with the consequence that sources in which the flaring mode
occurs, such as GRS\,1915+105 and
Cyg\,X-3, have radio power spectra that lie
more than an order of magnitude above the corresponding X-ray power
spectra. Of the four microquasars for which we have examined data, in only
one, Cyg\,X-1, is the flaring mode seemingly inactive.  While Cyg\,X-1 is a
black-hole candidate, one of the three flaring sources, Sco\,X-1, is a
neutron star. Consequently, it is likely that both modes are driven by the
accretion disk rather than black-hole spin. Radio imaging strongly suggests
that the flaring mode involves relativistic jets.

A typical microquasar is in the flaring mode a few percent of the
time, which is similar to the fraction of quasars that are radio
loud. Thus there may be no essential difference between radio-loud and
radio-quiet quasars; radio loudness may simply be a function of the
epoch at which the source is observed.

\end{abstract}


\section{Introduction}
\label{sec:intro}

Accreting stellar-mass black holes appear in a number of respects to
be scale models of the accreting supermassive black holes that power
active galactic nuclei (AGN). In particular, these sources, which have
been dubbed microquasars \citep{Mir99}, display power-law non-thermal
spectra and jets that are not infrequently strongly relativistic
\citep[e.g.][]{Mir94,Hje95,Fen04b}.  The existence of microquasars
presumably indicates that the physics of accreting black holes is
dominated by classical gravity and electromagnetism, which are
entirely scale-free theories.  Whatever its theoretical significance,
the scaling from quasars to microquasars implies that the evolution of
a source in the Milky Way is similar to that of an AGN but speeded up
by a factor of order $10^7-10^8$ (the ratio of the masses of the black
holes). Hence by studying a stellar-mass object over a human lifetime
one can gain insights into the variability of AGN over time-scales in
excess of $1\Gyr$, and thus extrapolate from observed properties of
AGN to the long-term time-averaged properties of these objects.

\begin{figure*}
\centerline{\psfig{file=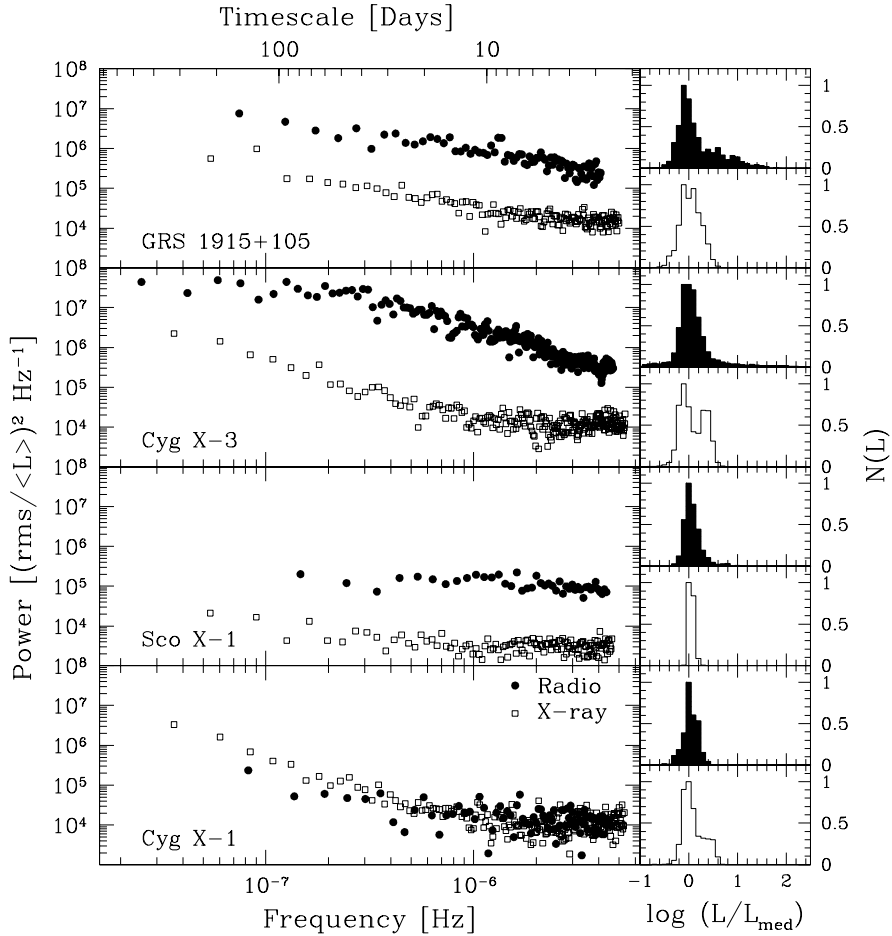,width=0.7\hsize}}
 \caption{The left panels show power spectra of GRS\,1915+105,
Cyg\,X-3, Sco\,X-1 and Cyg\,X-1 for the radio (filled) and X-ray
(open) data.  For each source the right panels show the distributions
of the daily-averaged radio (filled histogram) and X-ray (empty
histogram) data. \label{xrpsh}}
\end{figure*}

It is conventional to divide the most luminous AGN into radio-quiet
and radio-loud quasars \citep{Kel89,Mil90,Mil93,Kel94}.  Despite
considerable effort over several decades, it remains unclear what
determines the radio-loudness of a given quasar. Only a small minority
of quasars are radio-loud \citep{Ive02}. It is widely suspected that
large-scale radio emission of quasars is variable on time-scales of
$\sim10\Myr$ \citep{Sch00}, and a possibility that has often been
discussed is that many quasars that are now radio-quiet are at times
radio-loud, while currently radio-loud quasars may often be
radio-quiet. It is then natural to associate different modes of radio
emission of microquasars with radio properties of AGN
\citep{Mei01,Gal03,Mac03,Mer03,Fal04}.

In this paper we use data on the X-ray and radio variability of
microquasars to identify the stellar-mass analogues of radio-loud
quasars, and demonstrate that the microquasars Cyg\,X-3 and
GRS\,1915+105 spend a few percent of their time in the radio-loud
mode.  By contrast Cyg\,X-1 shows no sign of entering the radio-loud
regime. We present data that suggest that two distinct physical modes
are at work in the radio emission of microquasars. One, which we call
the `coupled' mode, causes the X-ray and radio fluxes to vary
approximately together \citep{Gle04} and never causes the system to
become radio-loud. The other, which we call the `flaring' mode,
dramatically boosts the radio luminosity of the source, making it
radio-loud, by causing dramatic ejection events. These events have
little impact on the X-ray luminosity \citep[although see][]{Mir98}.

\section{Comparison of the variability power in X-rays and in radio
  for microquasars}
\label{sec:comparevariability}

This paper is based on radio and X-ray monitoring data of the X-ray
binaries GRS\,1915+105, Cyg\,X-3, Sco\,X-1 and Cyg\,X-1 taken
respectively from the archive of the US National Radio Astronomy
Observatory Green Bank
Interferometer\footnote{http://www.gb.nrao.edu/fgdocs/gbi/gbint.html}
(GBI) and the Rossi X-ray Timing Explorer All-Sky
Monitor\footnote{http://xte.mit.edu/ASM\_lc.html} (ASM).  In
particular, we consider daily-averaged 2-10 keV ASM light-curves and
daily-averaged 2.25 GHz GBI light-curves. The X-ray light-curves of
the four sources refer to the period $50087-53300$ (J.D.$-2400000$);
the periods covered by the radio light curves differ from source to
source ($49485-51823$ for GRS\,1915+105, $44915-51823$ for Cyg\,X-3,
$50638-51823$ for Sco\,X-1, $50412-51823$ for Cyg\,X-1;
J.D.$-2400000$).  We estimated the power-spectra of the unequally
spaced light curves by computing Lomb's periodgrams \citep{Lom76}, and
subtracting statistical noise due to observational uncertainties. We
adopt the usual normalization, so the power spectrum is in units of
$(\rms/\av{L})^2 Hz^{-1}$, where $\av{L}$ is the average luminosity.

The bottom left panel of Fig.\ \ref{xrpsh} shows for Cyg\,X-1 the power
spectrum of emission in the radio (filled points) and X-ray (open
points) bands. The power spectra have very similar normalizations and
there is at most a suggestion that the radio points define a slightly
flatter slope. Thus the power
spectrum of this source is the same in the radio and X-ray bands.  The
(filled) histogram of radio powers on the right indicates that Cyg\,X-1
is not highly variable.  The corresponding distribution of X-ray
luminosity (empty histogram) is somewhat wider than the radio
distribution.

The top panels in Fig.\ \ref{xrpsh} tell a very different story for
GRS\,1915+105.  The radio power spectrum still has a similar shape to
the X-ray one, but its normalization is higher by a factor
$\sim30$. The histogram of radio luminosities in the top right panel
reveals a hump of low-luminosity events that is very similar to that
seen in the histogram for Cyg\,X-1, but in the case of GRS\,1915+105
there is a long tail to high luminosity that is absent in the case of
Cyg\,X-1. It is this tail of high-luminosity radio flares that lifts
the radio power spectrum clear of the X-ray power spectrum of the same
object.

The second layer of panels in Fig.\ \ref{xrpsh} shows that Cyg\,X-3
has radio and X-ray statistics that mirror those of GRS\,1915+105
rather than those of Cyg\,X-1. The third layer of panels shows data
for the neutron-star powered source Sco\,X-1.  As in the cases of
GRS\,1915+105 and Cyg\,X-3, the radio power spectrum lies well above
the X-ray power spectrum, but the offset is now slightly smaller,
though still larger than an order of magnitude. The cause of the
offset is again a tail of events that have high radio luminosity. This
tail is not as prominent as in the top two histograms.

\begin{figure}
\psfig{file=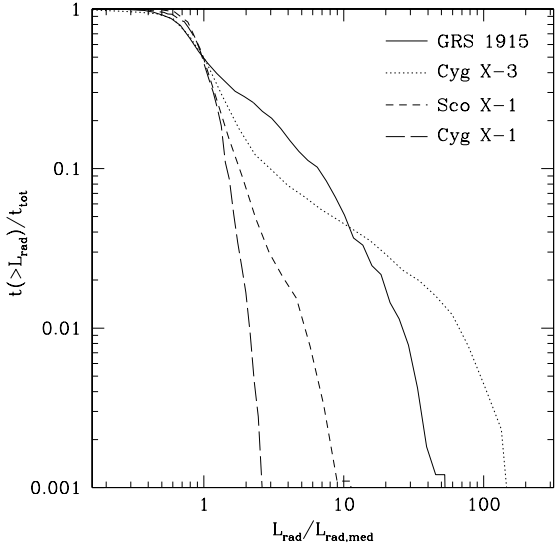,width=.9\hsize}
 \caption{Cumulative distribution of the daily-averaged radio data for the 
sources shown in Fig.\ \ref{xrpsh}.\label{cumul}}
\end{figure}

None of the X-ray histograms has a high-luminosity tail. The X-ray
histograms of the three black-hole candidates are all slightly wider
than the peaks in the corresponding radio histograms, and it is
possible that the underlying X-ray distributions are themselves
bimodal, especially in the case of Cyg\,X-3 and Cyg\,X-1. This is
consistent with the well known existence in these sources of two
distinct X-ray states (`low/hard' and `high/soft'). Sco\,X-1, unlike
the three black-hole candidates, has an X-ray histogram that is
distinctly narrower than the peaked part of the radio histogram.

The narrowness of the X-ray histogram for Sco\,X-1 may indicate that
the fluctuating X-ray luminosity from the accretion flow and jet is
superimposed on a steady contribution $L_{\rm NS}$ to the X-ray
luminosity from the surface of the neutron star \citep{Whi86}. Then at
a given frequency the normalized power spectrum of Sco\,X-1 in
Fig.~\ref{xrpsh} would be shifted downwards by a factor $(1+L_{\rm
NS}/\langle L_{\rm acc}\rangle)^2$, where $L_{\rm acc}(t)$ is the
fluctuating accretion luminosity.  The figure suggests that this
factor is $\sim4$, so we infer that in Sco\,X-1 about $\fracj12$ of
the mean X-ray luminosity comes from the surface of the neutron star.

\begin{figure}
\psfig{file=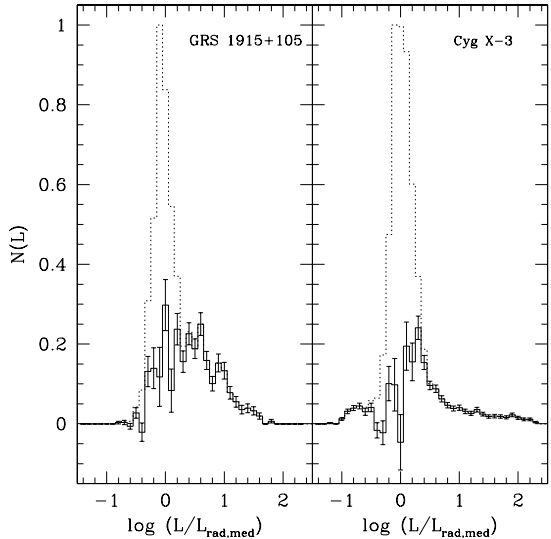,width=.9\hsize}
\caption{The solid line shows the result of subtracting the radio
histogram of Cyg\,X-1 from the radio histograms of GRS\,1915+105 and
Cyg\,X-3; the dotted line shows the original shapes of the
GRS\,1915+105 and Cyg\,X-3 distributions. Bars indicate Poissonian
uncertainties. \label{subh}}
\end{figure}

A natural interpretation of the data in Fig.\ \ref{xrpsh} is that,
apart from steady emission of the neutron star in Sco\,X-1, the X-ray
emission of all four sources is produced by the same, mildly
fluctuating, mechanism. This mechanism produces radio luminosity that
fluctuates with similar amplitude (although not necessarily with the
same phase) as the X-ray luminosity. On top of this baseline activity,
some strongly variable process contributes to the radio luminosity but
not to the X-ray luminosity.  A few percent of the time, the radio
luminosities of GRS\,1915+105 and Cyg\,X-3 are completely dominated by
this second process. The radio luminosity of Sco\,X-1 is more rarely
dominated by the second process, while the radio luminosity of Cyg\,X-1
never is. This difference in the radio properties of the sources is
apparent from Fig.\ \ref{cumul}, which shows the fraction of the time
each source spends above any given radio luminosity.  There is only a
one percent probability of finding Cyg\,X-1 more luminous than twice
its median luminosity, while there is about the same chance of
catching Cyg\,X-3 with a radio luminosity 70 times the median value.

Fig.\ \ref{subh} reinforces the picture that two distinct modes 
contribute to radio activity by showing that most of the
cluster of low-luminosity events in GRS\,1915+105 can be accounted for by
subtracting the histogram of radio events in Cyg\,X-1 after shifting it
horizontally and scaling it vertically but not horizontally. A similar
subtraction of the data for Cyg\,X-1 works only slightly less well in
the case of Cyg\,X-3. Thus all three black-hole candidates have
extremely similar distributions of low-luminosity radio events.

The three microquasars that have strong radio variability are observed
to have intermittent relativistic jets: GRS\,1915+105 \citep{Mir94},
Cyg\,X-3 \citep{Mio01,Mil04} and Sco\,X-1 \citep{Fom01}.  On the other
hand, there is no compelling evidence that Cyg\,X-1 has ever produced
relativistic jets: some elongated radio features are plausibly
interpreted as indicative of outflow \citep{Sti01}, but existing data
do not warrant the claim that it is relativistic.  Certainly, flares
in the radio light curve of Cyg\,X-1 are extremely rare
\citep{Poo04}. \cite{Fen04a} concluded that in jet-producing
microquasars, the mechanical luminosity of the jets probably dominates
the total energy output of the system regardless of the system's
state. Thus changes in the system's luminosity are expected to be
closely coupled to changes in the rate at which energy is carried
outwards by jets.  Moreover, episodes of enhanced radio luminosity are
quickly followed by episodes in which fast-moving knots of emission
are seen in radio images \citep{Mil04}.

\section{Implications for quasars}
\label{sec:implquasars}

While X-ray emission from quasars is invariably a core phenomenon,
radio emission from quasars sometimes extends over many kiloparsecs
and involves extended radio lobes, that are generally agreed to be
powered by relativistic jets. We think it is helpful to define
radio-loud quasars to be those with extended radio emission, and
radio-quiet quasars to be those with at most compact radio
emission. Since X-ray emission is invariably a core phenomenon, it can
be closely coupled with total radio luminosity only in radio-quiet
quasars. This consideration strongly suggests a parallel between
radio-loudness in quasars and the flaring mode in microquasars, and
hence an association of radio-quiet quasars with non-flaring states of
microquasars (either low/hard or high/soft).  This is in agreement
with the proposed association of radio-quiet AGN with high/soft
states \citep{Mac03}, but it conflicts with the suggested association
\citep{Fal04} of FR\,I radio sources with low/hard states.

We suggest that the radio emission in quasars that is associated with
the coupled mode, is confined to what is referred to as the core
emission from quasars, and is most readily detectable at GHz
frequencies.  Indeed, this core emission is observed from many quasars
classified as radio-quiet \citep{Blu98}.  We suggest that the
`flaring' mode leads to the formation of the large-scale ($>10\,{\rm
kpc}$) jets that are the hallmark of radio-loud quasars, be they
FR\,Is or FR\,IIs.

The radio luminosity to which jets give rise may be proportionally
smaller in microquasars than in radio galaxies because the ambient
medium around microquasars may not provide a working surface on which
a jet can randomize its bulk kinetic energy to produce hot spots, as
happens in the case of a radio galaxy \citep{Hei02}. Thus the
near-perfect scaling of black hole accretion from microquasar to
quasar is broken by a failure of the environment to scale in the same
way, with the result that equivalent enhancements in the jet power of
a microquasar and a quasar do not produce equivalent enhancements in
the radio luminosity.

This failure of perfect scaling should be borne in mind when comparing
the fraction of quasars that are radio-loud with the fraction of the
time that a typical microquasar is flaring. Specifically we should set
the threshold in radio luminosity at which we deem a microquasar to be
flaring at a smaller multiple of the median luminosity than the ratio
of the median radio luminosities of radio-loud and radio-quiet
quasars. The chosen threshold should be large enough that it excludes
all states in which the radio and X-ray luminosities are coupled. The
data for Cyg\,X-1, in which flaring is extremely rare \citep{Poo04},
together with Fig.~\ref{subh} which demonstrates the similarity in all
microquasars of the radio emission associated with the coupled mode,
suggests that a suitable threshold is three times the median
luminosity. The data of Fig.~\ref{cumul} show that GRS\,1915+105 is
over this threshold 21 percent of the time, while the corresponding
time fractions for Cyg\,X-3 and Sco\,X-1 are 10 and 3 percent,
respectively.  \cite{Ive02} find that 8 per cent of the bright
($i<18.5$) Sloan quasars are radio-loud in the sense that the ratio
$R$ of radio to X-ray flux exceeds unity.  In addition, the bottom
right panel of Fig. 2 in \cite{Ive04} shows that the distribution of
$R$ is bimodal. This distribution is qualitatively similar to the
distribution of the radio to X-ray flux from GRS\,1915+105 we show in
Fig.~1. In detail the distribution of $R$ in \cite{Ive04} is wider,
but this finding is consistent with the expected higher efficiency of
quasars in producing high radio luminosity.

In this picture, it is not surprising that the black-hole mass to
radio luminosity relation for AGN is observed to have a large scatter
at fixed black-hole mass \citep{Lac01,Mer03,Fal04}. Furthermore, the
radio luminosity function of AGN will be dominated by the effects of
variability \citep{Nip05}.

\section{Implications for physical models}
\label{sec:implmodels}

The presented data indicate that two distinct modes of energy output
are at work in microquasars: one producing relatively steady radio
emission coupled with X-ray emission, the other responsible for
strongly variable, flaring radio emission, apparently decoupled from
the X-ray emission. These two types of radio emission have been
referred to in the literature as `steady jets' and `transient jets',
respectively. Whether the physical circumstances of these phenomena
are the same as one another is still debated and is made less likely
by our findings. In particular it may be premature to assume that the
steady/coupled emission is generated by a jet as the transient/flaring
emission is.

Phenomenological studies yield information that may help us to answer
this question. Coupled radio emission is characteristic of the
low/hard X-ray state.  Flares often occur during a switch from the
low/hard to the high/soft state. Steady radio emission has a flat or
inverted radio spectrum and small Lorentz factors, while flaring
activity has often optically thin synchrotron spectrum and Lorentz
factors $>2$ \citep{Mil04,Fen05}.

From a theoretical point of view, there is a natural dichotomy in that
energy may be extracted either from the accretion disk, or from the
black hole's spin \citep{Bla77,Bla82,Pun90,Mei01}. The observation
that both the coupled and flaring mechanisms are active in the
neutron-star powered source Sco\,X-1 indicates that neither mechanism
is associated with the extraction of energy from black-hole spin, as
also pointed out by \cite{Fen04b} who observed an ultra-relativistic
outflow from the neutron-star microquasar Cir\,X-1.

In a plausible physical picture, the coupled mode comprises emission from a
dense non-thermal plasma at the base of a steady outflow from the system
\citep{Fen04a,Mar04}, while the flaring mode comprises emission from a
shocked relativistic outflow at significant distance from the binary. The
highly relativistic electrons produced in these shocks are much less likely
to produce X-ray photons by the inverse-Compton process than are electrons
at the base of the flow, because the ambient density of photons rapidly
diminishes with distance from the accretion disk.  Consequently, synchrotron
emission from a flare is associated with negligible inverse-Compton
radiation, and the flare is evident only at radio frequencies.

\section{Concluding remarks}
\label{sec:conclusions}

Histograms of the X-ray and radio power from four microquasars show
similar pronounced peaks at the low-power ends of the
distributions. In three of the four histograms of radio power, there
is a long tail to high luminosity that is lacking in the X-ray
histograms. We interpret this finding as indicating that two modes
are active in these systems. The `coupled' mode fluctuates only
mildly and produces both X-rays and radio power. The `flaring' 
mode is highly variable, and produces exclusively radio
power. Even though the flaring mode is active only a few percent of
the time, it typically dominates the mean radio-luminosity of the
system.

What we refer to in this paper as the coupled mode comprises, in the
picture of \cite{Fen04a}, radiation by a dense non-thermal plasma that
usually occupies a hole at the centre of the accretion disk, while the
flaring mode comprises synchrotron radiation from electrons that are
accelerated when the residue of this plasma is expelled at
relativistic speed as the hole in disk is intermittently filled
in. The absence of significant X-ray emission during these episodes is
natural if the electrons are shock accelerated far from the accretion
disk.

It seems likely that radio-quiet quasars are the analogues of
microquasars that are currently not flaring. This interpretation
implies that all radio-quiet quasars should display the radio emission
characteristic of the coupled mechanism. There are indications that
this is indeed the case, in that sensitive observations of radio-quiet
quasars do reveal core radio emission for example the sample of
Palomar-Green radio-quiet quasars observed with milli-arcsec
resolution by \cite{Blu98}.  Moreover, the fraction $f_{\rm Q}$ of
quasars that are radio-loud should be the same as the fraction of the
time in which a typical microquasar has a ratio of radio to X-ray flux
densities that is above an appropriate threshold. Results from the
SDSS \citep{Ive02} indicate that $f_{\rm Q}\approx0.08$, which is
consistent with the mean fraction of the time that the three flaring
microquasars are radio-flare-dominated. The apparent absence of
extended low-frequency radio lobes associated with radio-quiet quasars
(left-over from a putative previous radio-loud phase) does {\em not}
conflict with this picture, given the short radiative lifetimes of
synchrotron particles in extending lobes when all plasma loss
mechanisms are properly accounted for (Blundell \& Rawlings~2000).

\section*{Acknowledgments}

K.M.B.\ thanks the Royal Society for a University Research Fellowship.
The Green Bank Interferometer was a facility of the  National Science
Foundation operated by the National Radio Astronomy Observatory. 
This research has made use of Massachusetts Institute of Technology's
Rossi X-ray Timing Explorer All Sky Monitor.

\label{lastpage}
\end{document}